# Structures with Vertically Stacked Ge/Si Quantum Dots for Logical Operations

**Yu. N. Morokov**[a][^], **M. P. Fedoruk**[a], **A. V. Dvurechenskii**[b], **A. F. Zinov'eva**[b], and **A. V. Nenashev**[b]

[a]*Institute of Computational Technologies, Siberian Branch, Russian Academy of Sciences, pr. Akad. Lavrent'eva 6, Novosibirsk, 630090 Russia*
[^]*email: quant@ict.nsc.ru*
[b]*Rzhanov Institute of Semiconductor Physics, Siberian Branch, Russian Academy of Sciences, pr. Akad. Lavrent'eva 13, Novosibirsk, 630090 Russia*

**Abstract**—Ge/Si structures with vertically stacked quantum dots are simulated to implement the basic elements of a quantum computer for operation with electron spin states. Elastic-strain fields are simulated using the conjugate gradient method and an atomistic model based on the Keating potential. Calculations are performed in the cluster approximation using clusters containing about three million atoms belonging to 150 coordination spheres. The spatial distributions of the strain energy density and electron potential energy are calculated for different valleys forming the bottom of the silicon conduction band. It is shown that the development of multilayer structures with vertically stacked quantum dots makes it possible to fabricate deep potential wells for electrons with vertical tunnel coupling.

1. INTRODUCTION

Currently, both abroad and in Russia, hardware components for quantum calculations are being developed. Studies are being performed in several directions: the use of quantum states of individual particles in electromagnetic traps at low and ultralow temperatures, control of electromagnetic wave propagation in photonic crystals, implementation of solid-state-based qubits: superconducting qubits [1], NVcenters in diamond [2], phosphorus impurity atoms in silicon [3], and semiconductor quantum dots in which both the spin state [4] and the charge state [5] of a localized electron or hole can be an information carrier. This study is devoted to the search for opportunities for electron localization in Ge/Si multilayer structures with quantum dots for the purpose of using the localized states of carriers as qubits.

During the epitaxial growth of germanium on crystalline silicon, elastic strains develop at the heterointerface due to the different lattice constants of germanium and silicon. At the beginning of growth, a strained wetting layer is formed, which consists of several atomic germanium layers. In the course of further growth, pyramidal islands are formed on the wetting layer's surface. The key role in the self-assembly of the formed nanostructures is played by inhomogeneous elastic strains in the system [6, 7]. The formed islands several tens of nanometers in size are considered as quantum dots, since carriers localized in them exhibit a discrete spectrum.

As a rule, each study devoted to simulation of the electronic structure of self-assembled quantum dots starts with calculation of the elastic-strain distribution. This is due to the fact that the strain causes a change (distortion) in the band structure of the semiconductor materials; hence, has an effect on the shape of the potential well for electrons and holes in semiconductors. The largest number of studies on modeling the electronic structure and strain distribution in self-assembled quantum dots are devoted to the InGa/GaAs system [8, 9]; however, there are also some works on quantum dots in Ge/Si [9–11], InGaN/GaN [12], and InP [13] heterostructures.

In the Ge/Si heterostructure, holes are localized mostly in the bulk of quantum dots, in the germanium region. Trapped electrons are mostly localized in silicon, and potential wells for them are formed due to the elastic deformation of silicon layers surrounding the quantum dots [10], near the tops of the quantum dots and under their bases. Due to their deformations, maximum splitting of the sixfold degenerate Δ-valley of the conduction band occurs in these regions. Two valleys, $\Delta^{001}$ and $\Delta^{00\bar{1}}$, oriented along the [001] and [00$\bar{1}$] directions decrease in energy; the four others increase in energy. Localized electronic states are combined of the states of two bottom Δ-valleys.

During epitaxial growth, vertically stacked (in the growth direction) structures of germanium islands can be formed under certain conditions. The deformed field induced in silicon by a germanium island of the next layer creates favorable conditions for island creation in the next grown layer [14]. In such vertical structures, at certain distances between quantum dot layers, deformed fields are formed, and the potential well depth for electrons increases. Recently, this was demonstrated for a four-layer structure with vertically stacked quantum dots [15]. The energy of the electron localized in the central layer near the top of the quantum dot reached 60 meV.

The vertical structure of quantum dots formed during growth can be considered as a possible basic element for a quantum computer. Such a basic element for a quantum computer has a number of noticeable potential advantages in comparison with others.
(i) Spatial self-assembly of the vertical structure during growth.
(ii) Scale in the number of vertically ordered elements.
(iii) Possible real-time control of the parameters of each qubit in a technologically convenient way.
(iv) Large arrays of quantum dots with close parameters can be grown on each wetting layer. This makes it possible to consider large arrays of simultaneously operating computing elements.

As a particular example of a qubit, we can consider spin states of the electron localized near the quantum dot's top. However, one serious problem arises in such a consideration: there is a rather high potential barrier between neighboring vertically-stacked qubits, whose thickness is equal to the quantum dot height, which complicates solution of the problem of electron tunnel coupling. Possible electronic states corresponding to $\Delta^{100}$ and $\Delta^{\bar{1}00}$ valleys are free of this disadvantage. Such states are localized in silicon near edges on the quantum dot base; they are vertically separated by the minimum possible thickness of the layer of germanium atoms.

Let us consider the possibility of developing vertical structures of quantum dots, in which the electronic states localized in silicon near quantum dots will be mostly constructed of states of $\Delta^{100}$ and $\Delta^{\bar{1}00}$ valleys.

The first problem to be solved for practical operation with electrons of these valleys is to attempt to minimize the effect of competing $\Delta^{001}$ and $\Delta^{00\bar{1}}$ valleys deeper in energy. To solve this problem, we consider the case of the formation of vertical structures with the closest vertical arrangement of quantum dots, which excludes silicon interlayers in the region of the quantum dot tops. Such a situation is quite implementable experimentally, as shown in recent studies on the development of structures with vertically stacked quantum dots [16]. To be closer to the experiment, we will consider structures with partially overlapped volumes of pyramidal quantum dots. Such a structure makes it possible to implement a stack of truncated pyramids, which is, as a rule, observed in the experiment.

In this paper, we theoretically test the feasibility of the development of localized electronic states, tunnel coupled in the vertical, in Ge/Si heterostructures with vertically stacked quantum dots. To this end, we calculate the spatial distributions of the strain energy density and electron potential energy near quantum dots for different valleys that form the bottom of the silicon conduction band. The strain is calculated using the atomistic approach considering an actual diamond like lattice.

## 2. APPROACH DESCRIPTION

To simulate elastic-strain fields in quantum dots and in their surroundings, we previously used the discrete–continuous model [17] with the Keating potential. Some atoms in this model were considered in an explicit form; the effect of others was considered using Green's function calculated numerically. Unlike [17], in the present study we use a more accurate atomistic model and cluster approximation. Initially, all atoms are arranged at sites of the perfect diamond-like silicon lattice with the constant $a_{Si} = 0.543072$ nm which corresponds to the Si–Si bond length $l_{Si-Si} = 0.235137$ nm. In this case, the distance between neighboring atomic layers in the [001] direction is $b_{Si} = a_{Si}/4 = 0.135768$ nm. Substitution of individual silicon atoms with germanium atoms in this lattice results in local stresses in the structure due to different equilibrium bond lengths Si–Si, Si–Ge, and Ge–Ge. It is assumed that the subsequent elastic relaxation in the system conserves the topology of interatomic bonds of the diamond-like structure.

We consider single-layer and multilayer structures consisting of quantum dots arranged on wetting layers five atomic layers thick. The quantum dots are pyramids with a square base and a height-to-base size ratio of 1 : 10. The calculated clusters are constructed by a sequential increase in the number of coordination shells, beginning from a certain central atom. Basic calculations are performed for clusters containing atoms of 150 coordination shells; the central cluster atom is considered as belonging to the zeroth coordination sphere. The cluster of 150 coordination shells contains 2840951 atoms.

We use the following boundary conditions for the clusters. For atoms of two outer coordination shells (which also include atoms of the germanium wetting layer infinite in the $x$ and $y$ directions), the $x$- and $y$-coordinates are fixed, but the $z$-coordinates are completely free for the relaxation of all atoms (in the growth direction).

In the case of elastic relaxation of the system, the system's energy functional written as the Keating potential

$$E = \frac{3}{16}\sum_i\sum_j \frac{\alpha_{ij}}{l_{ij}^2} \cdot \left[(\mathbf{r}_i - \mathbf{r}_j)^2 - l_{ij}^2\right]^2 + \frac{3}{8}\sum_i\sum_{j>k}\frac{\beta_{ijk}}{l_{ij} \cdot l_{ik}} \cdot \left[(\mathbf{r}_i - \mathbf{r}_j)\cdot(\mathbf{r}_i - \mathbf{r}_k) + \frac{1}{3}l_{ij}\cdot l_{ik}\right]^2$$

is minimized [18], where $\mathbf{r}_i$ is the radius vector of the $i$-th atom: $\alpha_{ij}$, $\beta_{ijk}$, and $l_{ij}$ are the parameters depending on the atom type (the subscript $i$ enumerates all cluster atoms, and subscripts $j$ and $k$ enumerate the nearest neighbors of the $i$-th atom). The parameters $\alpha_{ij}$ and $\beta_{ijk}$ play the role of force constants, and $l_{ij}$ are the equilibrium bond lengths between atoms. The parameters of the potential are taken the same as in [17].

The total energy of the system is written as the sum of $E_i$ corresponding to contributions of individual atoms of the system,

$$E_i = \frac{3}{16}\sum_i \frac{\alpha_{ij}}{l_{ij}^2} \cdot \left[(\mathbf{r}_i - \mathbf{r}_j)^2 - l_{ij}^2\right]^2 + \frac{3}{8}\sum_{j>k}\frac{\beta_{ijk}}{l_{ij} \cdot l_{ik}} \cdot \left[(\mathbf{r}_i - \mathbf{r}_j)\cdot(\mathbf{r}_i - \mathbf{r}_k) + \frac{1}{3}l_{ij}\cdot l_{ik}\right]^2.$$

The quantity $E_i$ can be interpreted as the fraction of the elastic energy related to the $i$-th atom. The system energy is minimized using the conjugate gradient method. The numerical calculation is completed, when the change in the total cluster energy at one conjugate gradient step becomes smaller than the total cluster energy by 14 orders of magnitude. The accuracy limitation is caused by rounding errors when using double-precision numbers.

To calculate the single-particle quantum electronic states localized near the quantum dot, the effective electron potential energy can be written as the sum of the potential energy without regard for the lattice strain and the potential related to the elastic strain. If we use the effective-mass approximation and take the bottom of the conduction band as the reference point of electron energy, the electron potential energy in the strained silicon is reduced to the energy related to the lattice elastic strain.

Generally speaking, the lattice strain removes valley degeneracy, and the electron energy for different valleys will be given by [19]

for $\Delta^{100}$ and $\Delta^{\bar{1}00}$    $U_e = \Xi_d u(\mathbf{r}) + \Xi_u U_{xx}(\mathbf{r})$,
for $\Delta^{010}$ and $\Delta^{0\bar{1}0}$    $U_e = \Xi_d u(\mathbf{r}) + \Xi_u U_{yy}(\mathbf{r})$,
for $\Delta^{001}$ and $\Delta^{00\bar{1}}$    $U_e = \Xi_d u(\mathbf{r}) + \Xi_u U_{zz}(\mathbf{r})$,

Here, $\Xi_d$ and $\Xi_u$ are the strain potential constants, $U_{\alpha\beta}(\mathbf{r})$ is the strain tensor at the point $\mathbf{r}$, and $u(\mathbf{r}) = U_{xx}(\mathbf{r}) + U_{yy}(\mathbf{r}) + U_{zz}(\mathbf{r})$ is the strain tensor trace. We used the same strain potential constants for Si and Ge as in [19].

To calculate the strain tensor components, we used the following approach [17]. Let us consider the strain of a tetrahedron composed of the nearest neighbors of a certain lattice site. The tetrahedron's shape is defined by six parameters, e.g., lengths of its edges. Thus, the strain of this tetrahedron uniquely defines six components of the strain tensor related to this lattice site.

### 3. RESULTS

To test the used model, calculations were performed for single Ge/Si quantum dots of different sizes with a base half-width from 20 to 140 atomic layers. Figures 1–3 show the results for a single quantum dot with a base half-width of 100 atomic layers.

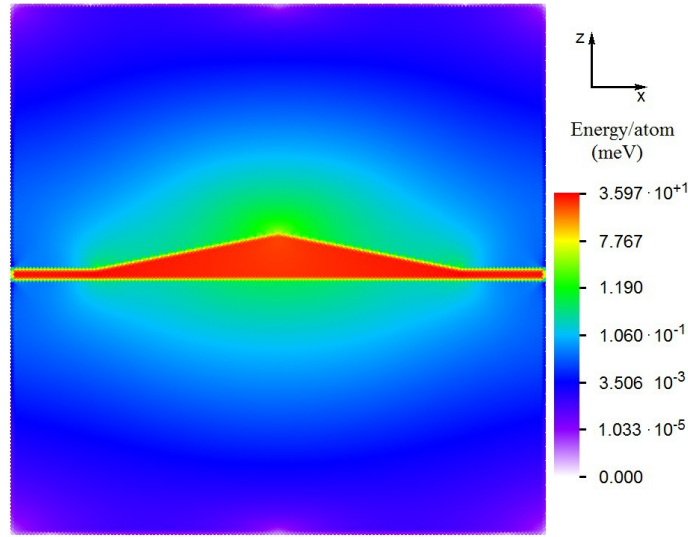

**Fig. 1.** Distribution of the bulk density of the strain energy in silicon in the central section of the cluster containing a single quantum dot with a base half-width of 100 atomic layers.

Figure 1 shows the distribution of the bulk density of the strain energy in silicon in the central section ($y = 0$) of the cluster. The choice of the section $y = 0$ is due to the fact that the largest values for the quantities under consideration are observed in this section in particular. We consider the distribution of the strain energy only in silicon, since distributions in germanium have been analyzed in sufficient detail previously [17] within the previous discrete–continuous model. According to our previous calculations [10], it was found that the potential wells for electrons in $\Delta^{001}$ and $\Delta^{00\bar{1}}$ silicon valleys are formed in the vicinity of the symmetry axis of the quantum dot, i.e., under the pyramid base and over its vertex. According to the cluster calculations performed, the depth of these potential wells is 349 meV (above the pyramid vertex) and 161 meV (below the pyramid base). The corresponding distribution of the electron potential energy density is shown in Fig. 2.

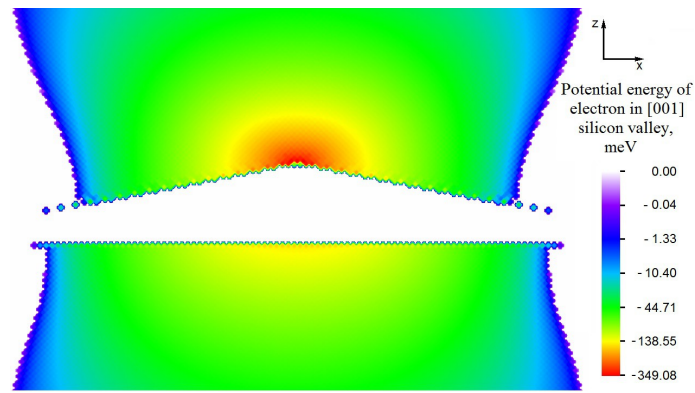

**Fig. 2.** Distribution of the electron potential energy in silicon for $\Delta^{001}$ and $\Delta^{00\bar{1}}$ valleys in the central section of the cluster containing a single quantum dot with a base half-width of 100 atomic layers.

Figure 3 shows the potential-energy distribution for the electronic states of $\Delta^{100}$ and $\Delta^{\bar{1}00}$ valleys. The depths of the potential wells formed near the quantum dot bases, above and below the wetting layer, are 124 and 62 meV, which is much smaller than the potential well depth of $\Delta^{001}$ and $\Delta^{00\bar{1}}$ valleys.

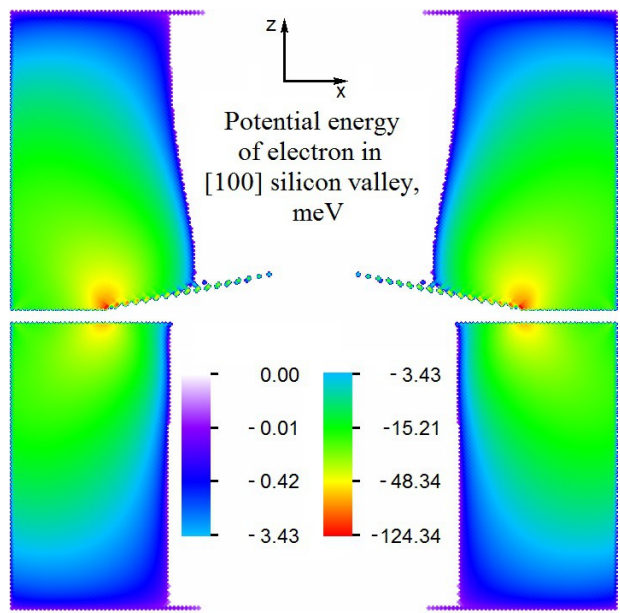

**Fig. 3.** Distribution of the electron potential energy in silicon for $\Delta^{001}$ and $\Delta^{00\bar{1}}$ valleys in the central section of the cluster containing a single quantum dot with a base half-width of 100 atomic layers.

We calculated the structures consisting of two vertically stacked quantum dots with a base half-width of 100 atomic layers. In this calculation series, the distance between the quantum dot centers was varied. The shortest distance was 16 atomic layers, which corresponds to a significant overlap of the pyramids. When quantum dots from large distances approach each other, the depth of the two outer potential wells (for $\Delta^{001}$ and $\Delta^{00\bar{1}}$ valleys) increases until the pyramids intersect. However, as the degree of pyramid overlap increases, the well depth again begins to decrease. In general, this is consistent with the problem of decreasing the effect of potential wells of this type. For the structure with a distance of 16 atomic layers between the centers of the overlapped pyramids, the depths of the two outer potential wells for the $\Delta^{001}$ and $\Delta^{00\bar{1}}$ and valleys are 182 meV (lower) and 410 meV

(upper). For $\Delta^{100}$ and $\Delta^{\bar{1}00}$ valleys, a maximum depth of 173 meV is reached above the bottom wetting layer. All these values exceed the corresponding depths of potential wells for the case of a single quantum dot.

For $\Delta^{100}$ and $\Delta^{\bar{1}00}$ valleys, the approach of quantum dots from large distances leads to monotonic deepening of the potential wells under consideration. This tendency is retained when pyramids intersect.

Thus, the formation of a structure with intersecting pyramids almost eliminates the potential wells for the $\Delta^{001}$ and $\Delta^{00\bar{1}}$ silicon valleys near the vertical axis of the system in the volume between the wetting layers. In the two outer regions, the potential wells for the $\Delta^{001}$ and $\Delta^{00\bar{1}}$ valleys remain and become deeper.

To decrease the deepest outer potential well for $\Delta^{001}$ and $\Delta^{00\bar{1}}$ valleys, which is formed above the top of the upper quantum dot, it is easiest to discard quantum dot growth on the uppermost wetting layer.

Calculations show that the depth of the potential well under the lower wetting layer is almost unchanged as the distance between wetting layers varies without pyramid cutoff. At the same time, a decrease in the distance in the case of pyramid cutoff rapidly decreases the depth of this potential well.

Thus, pyramid cutoff from above by a simple wetting layer (already without quantum dots) eliminates potential wells for the $\Delta^{001}$ and $\Delta^{00\bar{1}}$ valleys in the inner region between wetting layers and significantly decreases such wells in outer regions (below the lower wetting layer and above the upper wetting layer).

Let us now consider the vertical structure of nine equidistant wetting layers of equal thickness with the distance between the wetting layer centers equal to 16 atomic layers. On eight lower layers, identical quantum dots were grown with a base half-width of 100 atomic layers. Figure 4 shows the distribution of the electron potential energy in silicon for $\Delta^{001}$ and $\Delta^{00\bar{1}}$ valleys in the central section of the multilayer structure. The strain energies in silicon appreciably increase in all regions of the multilayer structure. This is due to an increase in the pressure of the increased array of germanium atoms and the more significant geometrical constraints for silicon atom relaxation in the inner regions of the structure under study. The maximum potential well depth for the electron in silicon for the $\Delta^{001}$ and $\Delta^{00\bar{1}}$ valleys under consideration in the central section ($y = 0$), shown in Fig. 4, is 318 meV. At the same time, the maximum depth of 411 meV for this cluster is reached for silicon atoms not in the $y = 0$ section, but in the lateral section ($z = $ const) corresponding to the second layer of silicon atoms under the fifth (from below) wetting layer. The deepest minima of the potential are on silicon atoms near vertices of the square of the upper cut of the fourth pyramid.

Despite the fact that the maximum depth of potential wells for $\Delta^{001}$ and $\Delta^{00\bar{1}}$ valleys is reached on central layers of the multilayer structure, it is unlikely that these wells can trap electrons, since they, as is seen in Fig. 4, have very small spatial dimensions, first of all, in the $z$ direction corresponding to these valleys. The presence of two spatially wide outer wells, i.e., above the upper wetting layer and below the lower wetting layer, seems to be more significant. These wells are seen in Fig. 4.

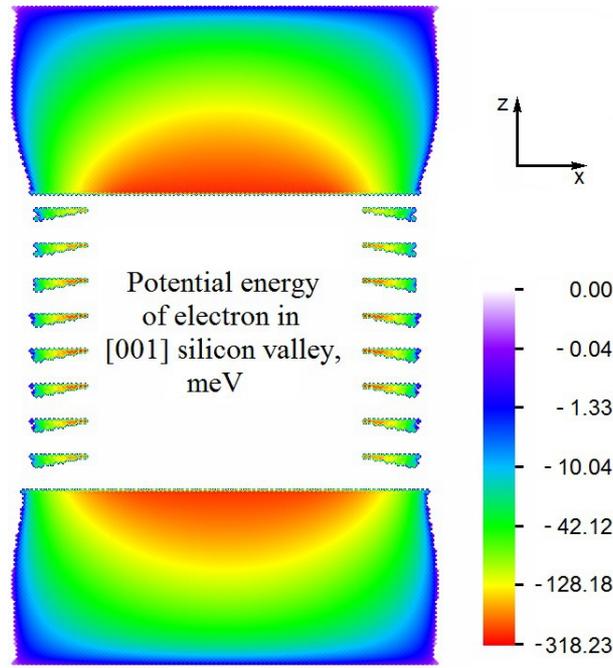

**Fig. 4.** Distribution of the electron potential energy in silicon for $\Delta^{001}$ and $\Delta^{00\bar{1}}$ valleys in the central section of the structure with eight quantum dot layers.

For both potential wells in Fig. 4, the largest depth is reached on the vertical axis of the system. The depth of the deeper upper well is 318 meV. The significant depth of the outer wells under consideration complicates the posed problem, i.e., suppression of the effect of the electronic states of for $\Delta^{001}$ and $\Delta^{00\bar{1}}$ valleys. The problem arises, either to use other methods for suppressing these outer wells or, on the contrary, to use in any way the presence of these wells in quantum algorithms.

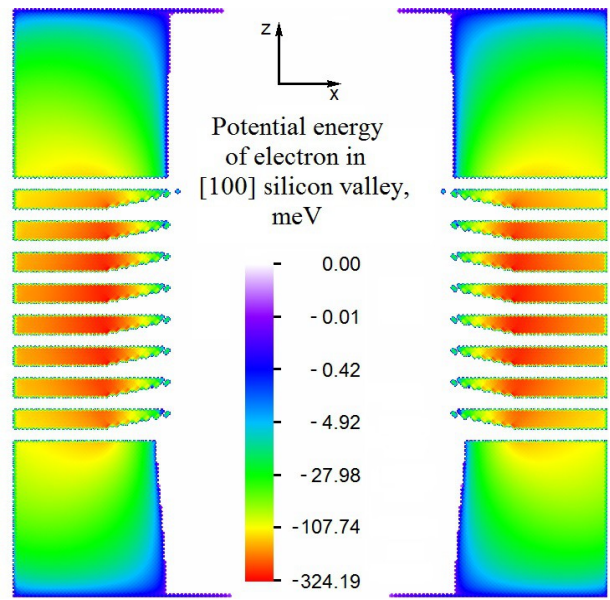

**Fig. 5.** Distribution of the electron potential energy in silicon for $\Delta^{100}$ and $\Delta^{\bar{1}00}$ valleys in the central section of the structure with eight quantum dot layers.

The distribution of the electron potential energy for $\Delta^{100}$ and $\Delta^{\bar{1}00}$ silicon valleys in the central section of the multilayer structure under consideration is shown in Fig. 5. The largest depth of potential wells for these valleys is 324 meV which is reached at $y = 0$ in the lateral section ($z =$ const) corresponding to the third layer of silicon atoms over the fifth wetting layer of the system.

Figure 6 shows the dependence of the lowest electron potential energy value on each lateral layer of the cluster under consideration in silicon for $\Delta^{100}$ and $\Delta^{\overline{1}00}$ valleys. The horizontal axis is the number of the cluster atomic layer along the $z$ vertical axis.

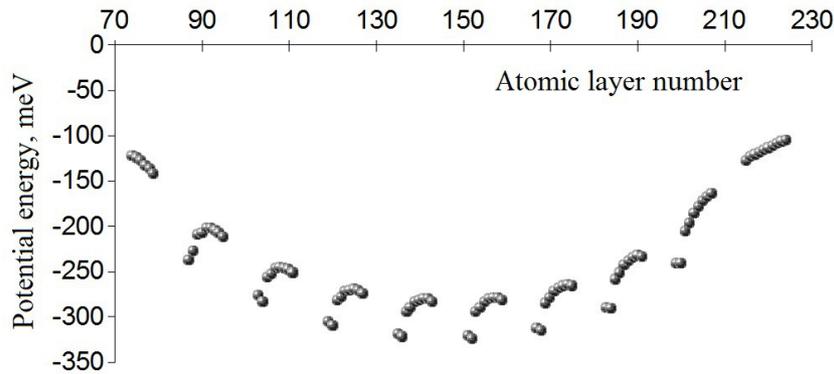

**Fig. 6.** Distribution of the electron potential energy in silicon for $\Delta^{100}$ and $\Delta^{\overline{1}00}$ valleys for the structure with eight quantum dot layers. Each point in the figure corresponds to the minimum energy on the corresponding lateral atomic layer of the cluster.

The table lists the energies (meV) corresponding to Fig. 6, being the lowest ones for each of the eight silicon layers of the structure under consideration.

| Silicon layer number | | | | | | | |
|---|---|---|---|---|---|---|---|
| 1 | 2 | 3 | 4 | 5 | 6 | 7 | 8 |
| -236.676 | -283.824 | -310.282 | -322.688 | -324.193 | -314.884 | -290.727 | -241.078 |

## 4. CONCLUSIONS

Ge/Si structures with vertically stacked quantum dots were simulated to implement solid-state basic elements of a quantum computer. The calculations performed showed the feasibility of the development of multilayer structures for operation with electrons in $\Delta^{100}$ and $\Delta^{\overline{1}00}$ valleys. For such structures, it is easier to organize controllable tunnel-coupling between the electrons of neighboring layers, which would make it possible to form series quantum registers for quantum calculations on their basis. However, the problem of the efficiency of electron localization in $\Delta^{100}$ and $\Delta^{\overline{1}00}$ valleys requires further investigation, including quantum-mechanical calculations.

## ACKNOWLEDGMENTS

This study was supported within the Integration project of the Siberian Branch of the Russian Acad emy of Sciences no. 43 "Development of the Physical Principles of Logic Gates Based on Quantum Dot Nanostructures".